\documentclass[aps,prl,amssymb,amsmath,showpacs,showkeys,reprint]{revtex4-1}

\newcommand{\BE}{\begin{equation}}
\newcommand{\EE}{\end{equation}}

\usepackage{color}
\usepackage{graphicx}
\usepackage{bm}

\begin{document}
\title{Nondiffusive suprathermal ion transport in simple magnetized toroidal plasmas}

\author{K. Gustafson}
\affiliation{Ecole Polytechnique F\'ed\'erale de Lausanne (EPFL),
Centre de Recherches en Physique des Plasmas, Association
Euratom-Conf\'ed\'eration Suisse, CH-1015 Lausanne, Switzerland}
\author{P. Ricci}
\affiliation{Ecole Polytechnique F\'ed\'erale de Lausanne (EPFL),
Centre de Recherches en Physique des Plasmas, Association
Euratom-Conf\'ed\'eration Suisse, CH-1015 Lausanne, Switzerland}
\author{I. Furno}
\affiliation{Ecole Polytechnique F\'ed\'erale de Lausanne (EPFL),
Centre de Recherches en Physique des Plasmas, Association
Euratom-Conf\'ed\'eration Suisse, CH-1015 Lausanne, Switzerland}
\author{A. Fasoli}
\affiliation{Ecole Polytechnique F\'ed\'erale de Lausanne (EPFL),
Centre de Recherches en Physique des Plasmas, Association
Euratom-Conf\'ed\'eration Suisse, CH-1015 Lausanne, Switzerland}
\date{\today}
\begin{abstract}
We investigate suprathermal ion dynamics in simple magnetized toroidal plasmas in the presence of
electrostatic turbulence driven by the ideal interchange instability.
Turbulent fields from fluid simulations are used in the non-relativistic equation of ion motion to compute 
suprathermal tracer ion trajectories. Suprathermal ion dispersion starts with a brief ballistic phase, during which 
particles do not interact with the plasma, followed by a turbulence interaction phase.  
In this one simple system, we observe the entire spectrum of suprathermal ion dynamics, from subdiffusion 
to superdiffusion, depending on beam energy and turbulence amplitude. We estimate the 
duration of the ballistic phase and identify basic mechanisms during the interaction phase 
that determine the character of suprathermal ion dispersion upon the beam energy and turbulence fluctuation 
amplitude.
\end{abstract}

\maketitle

Most plasmas are characterized by the presence of suprathermal particles, possibly generated by turbulent 
acceleration, external sources, or, as in the case of fusion devices, nuclear reactions. Understanding the 
basic phenomena that determine the suprathermal particle dynamics is a key challenge for the description of a 
wide range of plasma systems, ranging from magnetically confined plasmas for fusion 
\cite{HEIDBRINK:1994p178,Fasoli:2007p641,Gunter:2007,Albergante2011} to space plasmas \cite{Kaghash2006}.

In this Letter, we theoretically characterize suprathermal ion dynamics in the simple magnetized torus (SMT) 
configuration \cite{SMT}.  An SMT confines a plasma 
with a vertical magnetic field, $B_v$, superimposed on a toroidal magnetic field, $B_\phi$, creating helicoidal field 
lines that terminate on the vessel. This configuration is of interest to the plasma turbulence and fusion 
communities since it offers a simple and well-diagnosed testbed in which to study the basic physics of plasma 
turbulence and the associated transport of heat and particles, allowing parameter scans that are 
not possible in more complicated configurations.  In the SMT, turbulence has been characterized through global 
simulations \cite{Ricci:2010p620} validated against experimental data \cite{Ricci:2009p175,Ricci:2011p650}. 
These are unique simulations, which evolve
the plasma dynamics resulting from the interplay between the plasma source, losses at the vessel, and 
turbulence, with no separation between equilibrium and fluctuating quantities.  They provide the
turbulent fields for integrating suprathermal ion trajectories.

The SMT incorporates, in a simplified form, the fundamental elements determining suprathermal ion
dynamics, specifically: the Larmor gyration, the drifts related to the curvature and gradient of the magnetic field, the 
${\bf E} \times {\bf B}$ drift, and the polarization drift. The latter two are related to plasma turbulence and strongly 
dependent upon its topological properties. The relative simplicity of the SMT allows comprehensive 
quantification of the interplay between these phenomena. Since the key elements are the same, the framework 
established here can be applied for interpreting suprathermal ion dynamics in more complicated and diverse contexts.
Examples include fusion devices with high energy neutral beams and $\alpha$-particle production,
cosmic ray propagation, and solar wind interaction with the magnetosphere.

The primary diagnostic for studying the dispersion of suprathermal ions is the variance 
$\sigma_R^2(t) = \left<\delta R^2\right> \sim t^{\gamma_{_R}}$ of their radial displacements, 
$\delta R \equiv R( t) - R( 0)$, where $<>$ is an ensemble average over many particle trajectories. By numerically 
integrating the trajectories of suprathermal ions in simulated SMT turbulent fields, and by exploring wide ranges of 
particle energy and turbulence amplitude, we show that the ions have a complex motion, which in general 
cannot be considered diffusive. Our simulations show that suprathermal ion dispersion starts with a brief ballistic 
phase, during which particles do not interact with the plasma, resulting in $\gamma_R \simeq 2$. This phase is 
followed by a turbulence interaction phase, which surprisingly shows the entire spectrum of suprathermal
ion spreading: superdiffusive ($\gamma_R > 1$), 
diffusive ($\gamma_R = 1$),
or subdiffusive ($\gamma_R < 1$), depending on particle energy and turbulence amplitude. We provide
an estimate of the duration of the ballistic phase and we identify the mechanisms that determine, in the interaction phase, 
the dependence of $\gamma_R$ upon the beam energy and turbulence fluctuation amplitude.

For an SMT example, we refer to the parameters of the TORPEX device \cite{fasoli2006pop,fasolieps2010}, 
namely $B_v \ll B_{\phi}$, $\beta \ll 1$ and $T_i \ll T_e$. In TORPEX, a localized source of plasma on the high-field 
side of the torus is generated by microwave absorption at the electron-cyclotron and upper-hybrid resonances. 
A suprathermal ion beam is provided by a miniaturized lithium $6^+$ ion source \cite{Plyushchev:2006p460}. 
A number of turbulence regimes have been characterized 
both experimentally and theoretically \cite{Ricci:2010p620} for TORPEX. 
Here, we focus on the ideal interchange instability, which is 
dominant for a sufficiently high $B_v$ and low plasma resistivity. In this regime, $k_\parallel = 0$, and the 
wavelength of the dominant mode along $z$ (the direction perpendicular to both 
${\bf B}$ and the radial direction, which is tilted by an angle  $\theta = \tan^{-1}\left(B_v/B_{\phi}\right) \ll 1$ 
with respect to the vertical direction), is given by the return length 
of the field line on the poloidal plane, i.e. $\lambda_z=\Delta$, where $\Delta= 2 \pi R_0 B_v/B_\phi$, 
with $R_0$ being the SMT major radius. Since $k_\parallel =0 $, turbulence in this regime can be described 
by two-dimensional simulations in the tilted plane perpendicular to the magnetic field.

\begin{figure}[t]
\includegraphics[width=8cm]{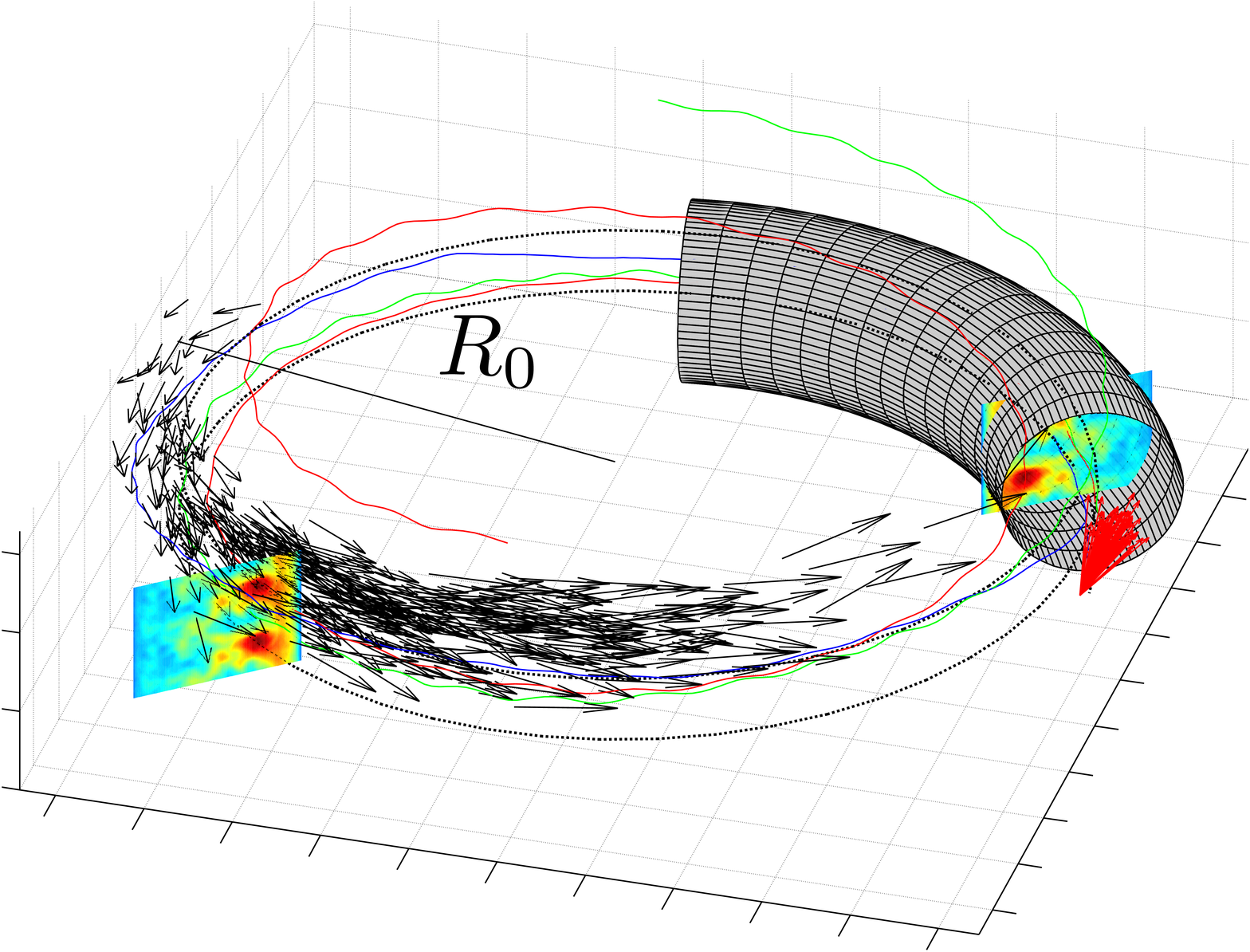}
\caption{(Color online) Suprathermal ions in a schematic of TORPEX. A section of
the torus vessel is indicated, as well as a magnetic field line (dashed black line).
Three suprathermal ion trajectories (red, green and blue solid lines) for a
$\mathcal{E} = 20$ injection are shown. The injection cone is indicated in red, 
while the black arrows indicate several ions at $t\Omega =130$.
A snapshot of the electrostatic potential, $\Phi$, obtained from the simulations in 
Ref.~\cite{Ricci:2009p175} and used to integrate the ion trajectories, is shown on two poloidal cross-sections.}
\label{fig:3dtraj}
\end{figure}

We integrate the non-relativistic ion equation of motion with the Lorentz force in the SMT magnetic 
configuration, considering the time-dependent electric field provided by the simulation described in 
Ref.~\cite{Ricci:2009p175}. 
An example of the simulated electrostatic potential for TORPEX, $\Phi$, is shown in 
Fig.~\ref{fig:3dtraj} in two perpendicular planes at different toroidal locations. Two distinct poloidal regions with 
different features of plasma turbulence have been identified previously in both simulations and experiments. 
First, a mode region exists on the high-field side of the SMT, 
where the plasma is generated and a coherent ideal interchange mode is present. Second, a region on the 
low-field side \cite{Furno:2008p95} with lower plasma density and temperature is marked by intermittent structures 
termed blobs \cite{theiler:055901,Ippolito2011}.

Injection conditions for the ions are chosen to mimic the suprathermal ion source in TORPEX, focusing on parallel 
injection, such that the axis of the injection cone is directed along a field line using a Gaussian angular distribution 
with a $0.1$rad variance. The initial velocities have a Gaussian distribution with a spread $\sigma_{v_0} = 0.1v_0$, 
with $v_0$ being the mean initial velocity, injected at the point $R=R_0$, a region where turbulence is 
transitioning from the mode region to the blob region. We consider the ions as tracer particles, such that they do 
not influence background fields, and ignore suprathermal ion collisions. These two assumptions are motivated 
by the experimental conditions in TORPEX.

We now detail the main elements determining a suprathermal ion trajectory in SMT turbulent fields. In the parallel direction, 
particle velocity is essentially unaffected since $k_\parallel = 0$. In the perpendicular plane, ion trajectories 
result approximately from the combined effects of four elements. These are the gyromotion, with Larmor radius 
$\rho = v_{\perp}/\Omega$ and Larmor frequency $\Omega=qB/m$ ($m$ and $q$ are the suprathermal ion mass 
and charge, respectively), the drifts related to the curvature and radial gradient of the magnetic field, 
$v_{\nabla \mathbf{B}}$, the $\mathbf{E}\times\mathbf{B}$ drift, and the polarization drift. 

The drift velocity $v_{\nabla \mathbf{B}} = (v_{\perp}^2/2 + v_\parallel^2 ) \mathbf{e}_z/(\Omega R)$ is oriented 
purely in the $z$ direction and dominated by the $v_\parallel^2$ curvature term for parallel 
injection. For the $\mathbf{E}\times\mathbf{B}$ drift, we note that the time-averaged 
electric field has a radial component that causes a drift in the $z$ direction. The fluctuating electric field leads to alternating 
displacements in the radial and $z$ directions. The size of these displacements is determined by the 
size and amplitude of the fluctuating vortex and blob-like structures in the turbulence, and by the Larmor radius. 
In fact, the $\mathbf{E}\times\mathbf{B}$ drift accounts for the gyroaveraged electric field. 
In the case where the suprathermal ion 
Larmor radius is significant compared to the scale of the turbulence, $k_{\Delta}\rho \gtrsim 1$ 
($k_\Delta=2 \pi /\Delta$), the gyroaverage decreases the magnitude of the $\mathbf{E}\times \mathbf{B}$ velocity 
with respect to cases for which  $k_{\Delta}\rho \ll 1$. Finally, the polarization drift causes modest but observable 
particle energization, similar to that seen in Ref.~\cite{Chandran:2010p556}. This can affect the transport slightly 
by increasing the average Larmor radius.

\begin{figure}[t]
\includegraphics[width=8cm]{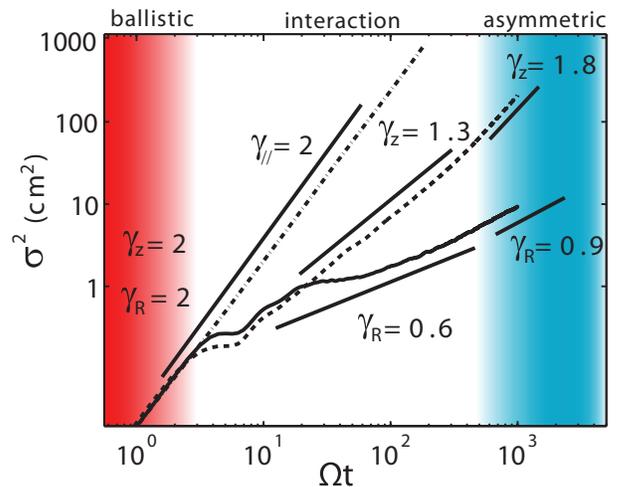}
\caption{(Color online) Variance of displacements in the radial (solid curve), vertical (dashed curve) 
and parallel (dash-dotted curve) directions for $\mathcal{E} = 50$ and $\xi = 0.8$ are shown. Dispersion exponents $\gamma$ are fitted with solid line segments, which have error of $\pm0.1$.  For the radial dispersion, an initial ballistic phase occurs (red-shaded region) with $\gamma_R \simeq 2$. This is followed by the turbulence interaction phase when $\gamma_R$ remains nearly constant. Later, the asymmetric phase (shaded blue region) shows an increased value of $\gamma_R$.  For the parallel direction, since there are no forces, $\gamma_\parallel \simeq 2$ always.  The $z$-directed spreading also
shows three phases in which the superdiffusion is due to $v_{\nabla\mathbf{B}}$.}
\label{fig:Gf4-torpex100}
\end{figure}

In the SMT, the turbulent $\mathbf{E}\times\mathbf{B}$ drift is thus the sole cause of suprathermal ion 
beam radial dispersion, which we quantify by the variance $\sigma_R^2(t)$ of the radial displacements. 
Figure \ref{fig:Gf4-torpex100} illustrates an example of the time evolution of $\sigma_R^2(t)$
with the three phases that appear in the simulations. These phases are characterized by different values 
of the $\gamma_R$ exponent, as determined by fitting a line in log-log plots of $\sigma^2(t)$. The phases can 
be categorized as ballistic, interaction and asymmetric, as shown in 
Fig.~\ref{fig:Gf4-torpex100}. They are distinct, but tend to transition smoothly 
from one into the next, as measured by the dispersion exponents.

The early ballistic phase is the relatively brief period with $ \gamma_R \sim 2$ before the ions interact significantly  
with the turbulence and magnetic field.  In fact, the simulations show that the ballistic phase lasts until 
$t \simeq \min \left\{2 / \Omega,m v_\perp / (q E) \right\}$. In principle, a gyrocenter ballistic 
phase may be present, during which the particle 
dispersion shows $\gamma_R \simeq 2$ superimposed on a Larmor oscillation. This gyrocenter ballistic phase 
would end when the gyrocenter speed changes significantly with respect to its initial value. In the present context, 
we estimate this phase to last less than one gyroperiod. The simulations agree that gyrocenter motion is never 
ballistic.

Following the ballistic phase, suprathermal ions interact with the plasma turbulence and the beam dispersion 
displays nearly constant $\gamma_R$. This quasi-steady state interaction phase ends when the ions have 
spread radially enough to sample regions where the local turbulent properties are significantly different than the 
ones at the injection position. At this time, the asymmetric phase begins because the
beam has spread to sample regions where the turbulence properties are significantly 
different with respect to the injection position.
Thus, the duration of the interaction phase is determined by the spread in the 
radial $\mathbf{E}\times\mathbf{B}$ velocity compared to the width of a roughly uniform region of turbulence.

\begin{figure}[t]
\includegraphics[width=8cm]{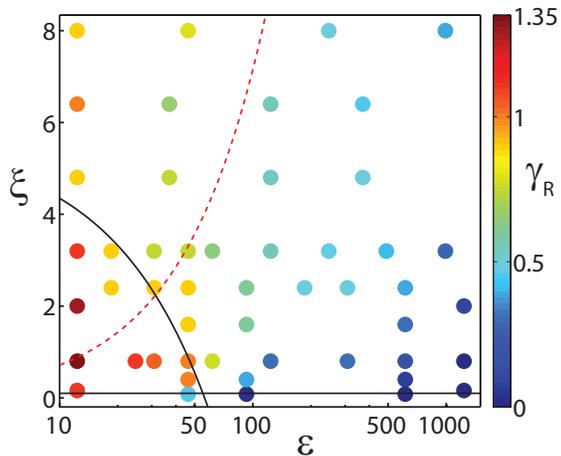}
\caption{(Color online) Dispersion exponents $\gamma_R$ (colored dots) in the interaction phase are displayed in the 
$\xi - \mathcal{E}$ plane. Error on the value of $\gamma_R$ is $\pm 0.1$. Regions for $\gamma_R$ are demarcated by the gyroaveraging condition ($k_\Delta\rho = 2$, solid black curve), by effective drift-averaging condition (Eq.~(\ref{eq:stepratio}), dashed red curve), and by the disconnected streamline condition ($\xi=0.3$, horizontal line).}
\label{fig:gamma_scatter}
\end{figure}

In Fig.~\ref{fig:gamma_scatter}, the values of $\gamma_R$ in the interaction phase, as obtained from a large 
number of simulations, are shown as a function of the beam energy, which we express in terms of 
$\mathcal{E} \equiv \mathcal{E}=mv_0^2/(2 T_e)$, 
and of the amplitude of the fluctuations of the electric field,  $\xi \equiv e\tilde{\Phi}/T_e$. 
Here, $T_e$ denotes the electron temperature, averaged over time and the $z$ 
coordinate, and $\tilde{\Phi}$ is the root mean square fluctuation amplitude of $\Phi$.  Both are 
evaluated at the injection point.  The fluctuation 
amplitude is scanned by rescaling the fluctuating part of the simulated electric field. Due to variations in the 
plasma properties, this is also equivalent to varying $\xi$ and $\mathcal{E}$ by injecting ions at different radial positions.  
At $R=R_0$, the typical experimental value is $\xi = 1.5$, while the simulations show smaller amplitude fluctuations, 
$\xi = 0.8$. Figure~\ref{fig:gamma_scatter} reveals the existence of different regimes for the ion dynamics. The 
complex dependence of $\gamma_R$ on these parameters is now explained.	

Superdiffusive dispersion, i.e. $1 < \gamma_R$, is observed in a region of the $\xi - \mathcal{E}$ plane, 
corresponding to $k_\perp\rho \to 0$, $v_{\nabla\mathbf{B}} \to 0$, and $\xi \gtrsim 0.3$. 
In this limit, turbulent structures are relatively static with respect to the ions, thus allowing ions to move large distances in a 
single direction \cite{DelCastilloNegrete:2005p426}. However, if the fluctuations are reduced below a certain
level, $\gamma_R$ drops dramatically because the amplitude of the vortex structures is too small for the 
structures to form connections between the center and edge of the plasma. This is a topological constraint set by 
the amplitude of the turbulent fluctuations. In Fig.~\ref{fig:gamma_scatter}, the $\xi \simeq 0.3$ boundary, for which 
connected velocity streamlines do not form, is indicated by the horizontal line: $\gamma_R$ decreases very 
sharply for $\xi \lesssim 0.3$.

The boundary marked by the solid curve in Fig.~\ref{fig:gamma_scatter} denote the points for which 
$k_\Delta \rho =2$. Outside this boundary gyroaveraging gradually decreases the number and the amplitude of the 
suprathermal ion radial displacements. This leads to a reduction of $\gamma_R$ from superdiffusive to a diffusive value, $\gamma_R \simeq 1$. 

The drift velocity $v_{\nabla \mathbf{B}}$ has an important effect on the suprathermal ion
dynamics. Provided that the motion along $z$ is sufficiently fast, an effective drift-average of the electric field
fluctuations reduces the radial dispersion, which is subdiffusive for a significant amount of time in
the interaction phase. Subdiffusive radial spreading occurs if the time required for an ion to traverse
a turbulent vortex along $z$ is significantly smaller than the time to traverse the same vortex radially.
This time can be estimated as follows. Let $\tau_R$ be the
time required to move radially across the structure, such that
\BE
\tau_R \sim L_R/v_{{\bf E}\times {\bf B},_R},
\EE
where $L_R$ is the radial extent of the vortex,
which has been estimated \cite{Ricci:2008p228} as $L_R \sim \sqrt{L_p/k_\Delta}$. Similarly, we define
\BE
\tau_z \sim L_z/\left(v_{\nabla \mathbf{B}}+ v_{{\bf E}\times {\bf B},z}\right)
\EE
as the time required to cross a vortex of size $L_z \sim 1/k_\Delta$ due to the velocity in
$z$. The curve
\begin{equation}
\label{eq:stepratio}
\frac{L_R \left(v_{\nabla \mathbf{B}}+ v_{{\bf E}\times {\bf B},z}\right)}{L_z v_{{\bf E}\times {\bf B},_R}} \simeq \chi
\end{equation}
approximately identifies the region where the ions are more likely to complete radial steps before drift-averaging makes radial steps less likely.
The numerical parameter $\chi \simeq 5$ is empirically observed to correspond to the subdiffusive transition
for all values of $\xi$ tested.
The condition for $v_{\nabla {\bf B}}$ averaging, given by Eq.~(\ref{eq:stepratio}), is also displayed in Fig.~\ref{fig:gamma_scatter}, confirming the reduction of the suprathermal ion dispersion rate to subdiffusive values due to drift-averaging.

Additional simulations with $R_0 \to \infty$ confirm the role of $v_{\nabla \mathbf{B}}$ drift in the suprathermal ion dispersion. 
In these simulations, dispersion is reduced from turbulent superdiffusion only by gyroaveraging. The values of 
$\gamma_R$ decrease from $\gamma_R > 1$ to $\gamma_R \simeq 1$ when the Larmor radius becomes 
sufficiently large, but the dispersion never becomes subdiffusive.

Remarkably, we find in one simple system that suprathermal ion spreading can be subdiffusive, diffusive, or 
superdiffusive depending on the ion energy and turbulence amplitude. In previous works, superdiffusion and subdiffusion,
separately, have been used to model plasmas, see 
e.~g.~\cite{PhysRevLett.40.38,DelCastilloNegrete:1998p269,Hauff07b,Sanchez:2008p471}.  The coexistence of the three 
regimes in an \textit{ad hoc} Hamiltonian model was
observed in Ref.~\cite{Abdullaev2000}.  Nevertheless, a diffusive approach continues to be assumed typically. 
Diffusion may sometimes describe suprathermal ion transport over short 
time and spatial scales, since a nondiffusive process may be linearized as an ``effective diffusion.''  Our 
simulations show that this approximation is valid only locally, since $\gamma_R$ can be drastically different than unity, 
and time-dependent. Therefore, the effective local suprathermal ion diffusivities can show a strong time dependence. In the 
cases explored here, they can be two orders of magnitude away from measurements of thermal particle diffusivity 
computed for the ideal interchange mode found in Ref.~\cite{Ricci:2009p428}.

Available data from the TORPEX device indicate that the magnitude of suprathermal ion dispersion is consistent with 
simulations at a single point in time during the ballistic phase \cite{bovet-iaea2011}. Our estimates suggest that the 
transition from ballistic to interaction-phase values of $\gamma_R$ should be observable for some experimental 
values of $\mathcal{E}$. However, the experimental measurements of $\gamma_R$ made so far are not sufficient 
to confirm our theoretical prediction, since a high resolution in the toroidal direction is required, which will be 
possible to achieve with a toroidally moving source under construction. Measuring a change in $\gamma_R$ 
into the asymmetric phase will be difficult because most of the ion beam tends to exit the plasma before this phase 
is well resolved.

The interplay of fundamental phenomena such as gyromotion and curvature drift,
which determine the transport of suprathermal ions in the SMT, is also present in fusion confinement configurations.
For example, orbit-averaging for trapped particles in a tokamak is analogous to drift-averaging for suprathermal ions in
TORPEX. Also, radial constraint of transport due to zonal flows associated with ion-temperature gradient (ITG)
turbulence \cite{Rogers:2000} or other velocity-shearing mechanisms
may lead to similar subdiffusive tendencies as we find to be caused by
vertical drift in the SMT.   Our results cover $10 \lesssim \mathcal{E} \lesssim 1000$, $k_\perp\rho \ll 1$ to $k_\perp\rho \sim 10$,
and Kubo numbers \cite{Kubo:1963,Hauff:2008p211} in the range $0.2 < \mathcal{K} \equiv v_\perp\tau_c/\lambda_c < 5$.  Here, $\tau_c$ and $\lambda_c$ are the correlation time and correlation length of the turbulence, respectively.  These ranges are relevant for present fusion devices with neutral beam 
injection \cite{Gunter:2007} and may be relevant for a future DEMO tokamak as well. 
While the SMT includes much of the fundamental physics for fusion plasmas, 
the general rules found here are relevant in other contexts as well, such as 
cosmic rays and solar flares \cite{1966ApJ146480J,HauffShalchi2010}.  

This work was supported by NSF-IRFP Award \#0853498 and by the Swiss National Science Foundation. We acknowledge useful discussions with A. Bovet, J. Graves, F. Halpern, J. Loizu and C. Theiler.

\end{document}